# Time-Resolved and Temperature Tuneable Measurements of Fluorescent Intensity using a Smartphone Fluorimeter

Md Arafat Hossain,[a,b] John Canning,[a,c,d]† Zhikang Yu,[a] Sandra Ast,[d] Peter J. Rutledge,[c] Joseph K.-H. Wong,[c] Abbas Jamalipour,[b] and Maxwell J. Crossley[c]

[a.] interdisciplinary Photonics Laboratories, School of Chemistry, The University of Sydney, NSW 2006, & School of Computing and Communications, University of Technology Sydney, NSW 2007, Australia.
[b.] School of Electrical and Information Engineering, The University of Sydney, NSW 2006, Australia.
[c.] School of Chemistry, The University of Sydney, NSW 2006, Australia.
[d.] Australian Sensing and Identification (AusSI) Systems, Sydney, NSW 2000, Australia.
† Corresponding author: john.canning@sydney.edu.au

A smartphone fluorimeter capable of time-based fluorescence intensity measurements at various temperatures is reported. Excitation is provided by an integrated UV LED ($\lambda_{ex}$ = 370 nm) and detection obtained using the in-built CMOS camera. A Peltier is integrated to allow measurements of the intensity over $T$ = 10 to 40 °C with a maximum temperature resolution of $\delta T$ ~ 0.1 °C. All components are controlled using a smartphone battery powered Arduino microcontroller and a customised Android application that allows sequential fluorescence imaging and quantification every $\delta t$ = 4 seconds. The temperature dependence of fluorescence intensity for four emitters (Rhodamine B, Rhodamine 6G, 5,10,15,20-tetraphenylporphyrin and 6-(1,4,8,11-tetraazacyclotetradecane)2-ethyl-naphthalimide) are characterised. The normalised fluorescence intensity over time of the latter chemosensor dye complex in the presence of $Zn^{2+}$ is observed to accelerate with an increasing rate constant, $k$ = 1.94 min$^{-1}$ at $T$ = 15 °C and $k$ = 3.64 min$^{-1}$ at $T$ = 30 °C, approaching a factor of ~ 2 with only a change in temperature of $\Delta T$ = 15 °C. Thermally tuning these twist and bend associated rates to optimise sensor approaches and device applications is proposed.

## Introduction

Smart device-based scientific instrumentation is democratizing scientific research and impacting fields such as medical, chemical and agricultural diagnostics as well as environmental and industrial monitoring.[1] Smart devices, such as smartphones, can be interconnected through wireless networking underpinning smart sensing networks and sensor smartgrids from which much of the so-called Internet-of-Things (IoT) derives. The majority of these instruments initially performed basic imaging and color applications such as microscopy[2,3] and colorimetry-based analysis.[4-7] The latter, in particular, has been important for analysis in the field chemical and biological samples. For example, colorimetric instrumentations using smartphone camera and external illumination has been reported repeatedly for analysing chlorine,[8] bromide ion,[9] *E. Coli* bacteria[10,11] and metal ions[12,13] in water samples. Recently, we introduced the concept of using a self-contained smartphone device[14] by development of a field-portable water quality monitoring instrument for pH measurements and real-time mapping. This device uses optical sources powered by the smartphone.[15,16] Smartphone colorimetric detections are reasonably integrated with other media such as micro-fluidic devices with potential bio-markers in many point-of-care diagnoses.[17]

The introduction of spectroscopy led to the development of a range of portable, field-worthy absorption and fluorescence spectrometers[18-23] and associated lab-in-a-phone technologies[14]. The acquisition of fluorescence spectra is particularly noteworthy given the significant number of biomedical applications that can benefit from a portable, lab-in-a-phone platform.[24] Although a great deal of work has been done towards steady-state fluorescence measurements on the smartphone platform,[14,16,18,24,25] there have been no previous reports on time-based measurements. These would offer ways to resolve the evolution of fundamental information and the relationships between different processes at a molecular level in chemical and biological science. Further, to be able to do this outside the laboratory and in the field would be a major advance in sensing.

The time scale of various chemical processes at a molecular level can range from femtosecond to millisecond and longer.[26] Longer processes are characterised by changes in physical parameters, often involving temperature. For example, the measurement over time of growing fluorescence at different temperatures of a naphthalimide chemosensor dye in metal ion complexes has been resolved, revealing features and processes characteristic of an optical molecular diode.[27] The mechanical movement associated with long timescale relaxation with optical excitation has even been proposed as the basis for molecular nanobot technology.[28] Further, by tuning the temperature of the sample, the molecule can reverse the selectivity between different metal ions having opposite redox potential, enabling a novel and low-cost method to discriminate between metal ions that are otherwise hard to distinguish.[27,29] These markers have potential applications in both biological and environmental diagnostics.

Generally, temperature ($T$) dependent fluorescence measurements are extremely useful for studying many other diagnostic processes in detail. Monitoring in-channel fluid temperatures using fluorescence in various microfluidic systems has been reported using temperature-dependent fluorescent dyes within a T-shaped microchannel intersection during electrokinetic pumping.[28,29] $T$-dependent fluorescence has been used to investigate temperature induced risks for hyperthermic stress or cell damage.[30] Low temperature fluorescence with gas chromatography is used to measure metal ions at ultra-low concentrations.[31] $T$ measurements can also be used to characterise molecular probes with regard to their intrinsic fluorescence switching properties and the thermal stability of different fluorescent sensor materials.[32] However, many of these examples have been characterised in laboratory settings because field-worthy instrumentation capable of time-based, temperature measurements of fluorescence are not readily available. Existing systems are connected to an external heating and cooling unit and are designed for benchtop operation, often with significant consumption of power that requires an external supply. The key challenge for achieving field-portable temperature measurements is the availability of a suitable, low-powered easy-to-control thermal unit. In previous work introducing the concept of democratisation of research through easily accessible and affordable field enabled instrumentation, we have reported the integration of photonic/optical components, such as ultra-violet (UV) light-emitting diode (LEDs) as the excitation source, with low power consumption into smart device circuits.[18] Here, we further add a Peltier heating element within a 3D-printed package designed to improve insulation and help mitigate power consumption and demonstrate a fully self-contained smartphone fluorimeter capable of both time-based and steady state measurements at different temperatures for the first time. An Arduino board plugged into the smartphone and controlled by an Android application allows the collection and internet-transmission of all data. The instrument is applicable for engaging a wide range of chemical researches, based on time and temperature based fluorescence, onto a field-portable smart device while serving global IoT connectivity via the internet.

Herein, the concept of time-resolved and temperature tuneable fluorescence measurements using an Arduino controlled smartphone fluorimeter is demonstrated on four systems: two laser dyes, a porphyrin emitter, and a metal ion chemosensor.

## Materials and methods

The complete system diagram of the time-resolved smartphone fluorimeter with a $T$ controlling unit is shown in Fig. 1. The fluorimeter contains an external microcontroller hardware interfaced with a smartphone, which holds all electronic components including the fluorescence excitation source, sample heating element and temperature sensing circuit. These components were operated via an Arduino Uno microcontroller module (ATmega328P–assembled), which is an open-source hardware and software prototyping platform, for building digital devices and interactive objects that can sense and control physical devices.[33] Such modules can send and receive commands either through wired connections or wirelessly, and are able to run multiple input-output devices simultaneously through its 14 digital output and 6 analog input ports. For instance, the fluorescence excitation source is connected to one of the digital output ports. ON-OFF switching and brightness of the excitation source can be controlled by tuning the voltage from the digital output port. Since the absorbance of most fluorescent sensor dyes (including our previously reported $Zn^{2+}$-responsive chemosensor dye[27]) lie in the UV region of the spectrum, a UV LED ($V$ = 3.0 V, $I$ = 20 mA, $\lambda_{ex}$ = 370 nm) was used as the fluorescence excitation source in this study. However, an excitation source at any other wavelength can also be added to the system, extending the fluorimeter's capabilities further. To raise or lower $T$ of the sample, a Peltier unit is used. Peltier devices are popular thermoelectric energy conversion modules used for heating or cooling any object/sample by applying potential across the junction of two different materials transferring energy from one side to the other (typically junctions of $P$-type and $N$-type semiconductors).[34] When one side of the block heats up the other side cools down and vice versa. A properly designed thermal isolation system can help to maintain the $T$ difference between hot and cold surfaces. Furthermore, to stabilize the temperature, a feedback loop is created by measuring the sample's $T$, feeding the data to the Peltier circuit and adjusting the current to deliver and maintain the desired $T$ and stability over time. The large surface area ($A$ = 2 x 2 cm$^2$) of the Peltier used in this work is sufficient to cover the surface of one side of the sample cuvette, enabling rapid heat transfer between Peltier and sample. By using a polarity reversing switch, the Peltier bias can be altered automatically. In the Peltier circuit, an $N$-type metal-oxide semiconductor field-effect transistor ($N$-MOSFET, $\alpha$ = 20) current amplifier circuit is used to amplify current to circumvent maximum currents from any output port ($I_{max}$ = 20 mA) of the Arduino and drive the Peltier. Thermal isolation between the Peltier surfaces was managed using an aluminum heat sink with a mounted cooling fan ($V$ = 3 to 5 V). Actual temperatures at the cuvette were measured using an infra-red (IR) temperature sensor with a range -(33 ± 0.6) ≤ T ≤ +(220 ± 0.6) °C. The Arduino board is connected and powered by the smartphone through the micro-USB port and data transfer is through an attached HC-06 Bluetooth module.

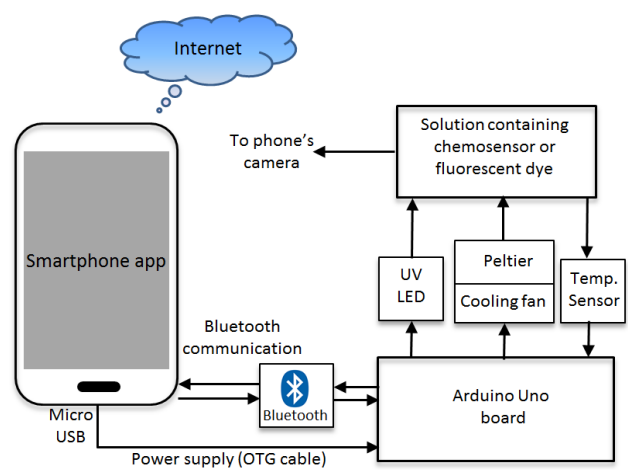

Fig. 1.  Layout of an Arduino microcontroller driven smartphone fluorimeter system.

A custom smartphone application software (app) analyses the fluorescence of the sample whilst controlling the devices and components on the Arduino board through Bluetooth communications. The app allows time-resolved as well as steady-state fluorescence detection on the smartphone camera. The temporal resolution is limited by the camera recording time to a few seconds and can be improved by replacing the imaging camera with a photodetector. For the applications reported here, the dominant relaxation time of the chemosensor dyes that will affect real field measurements arises from mechanical relaxation on a minute time scale so high temporal resolution is not required. Here, the smartphone app automatically captures images of the fluorescent sample at a specified time interval $\delta t$ (~ 4s), which is defined by the total recording time ($t$) and number of images ($n$) set by the user's input. In order to avoid other apps running simultaneously within the phone that impact processor speed and time, $\delta t$ is calibrated instead against a stop-watch. Once the image recording is completed, another command from the app determines the fluorescence intensity $F$, from the saved images, normalised against the initial fluorescence, $F_0$, from the stored images, if necessary and plots intensity over time as $F$ vs $t$. The algorithm to calculate fluorescence intensity from the captured images is reported elsewhere.[14,15] The smartphone fluorimeter can share the results of these measurements with other devices through the cloud and in principle relay this information anywhere. The data can be accompanied by location identification through GPS positioning taken automatically by the phone.

### 3D design, fabrication and packaging

The device casing was designed in AutoCAD and fabricated in-house with a 3D printer using acrylonitrile butadiene styrene (ABS) filament (Fig. 2). The final 3D design of the smartphone intensity fluorimeter consists of four different parts: an electronics panel, the sample chamber, the thermal unit, and the fluorimeter box. The fluorimeter box is a support frame to hold and align the smartphone at a 30° inclined to the horizontal axis for easier reading and optimises contact between the sample cuvette and the Peltier surface. On the top of the box where a frame holds the smartphone, a suitable port allows connection of the OTG cable from the smartphones micro-USB port to the power supply port of the Arduino – this can be packaged internally. Ventilation allows heat dissipation from all active components. Figure 2 (inset) shows the design of the sample chamber and the corresponding manufactured device. In order to reduce the levels of UV radiation reaching the CMOS detector directly and maximize the fluorescence readout, the UV LED is positioned orthogonal to the fluorescence emission path from the sample contained in standard quartz cuvette. An emission filter at $\lambda_{em}$ ~ 450 nm is used to further reduce background scattering which can be replaced or complemented by other bandpass filters where required. Access to the cell chamber for a standard cuvette is from the top whilst the temperature of the cuvette is monitored by the IR sensor from the bottom.

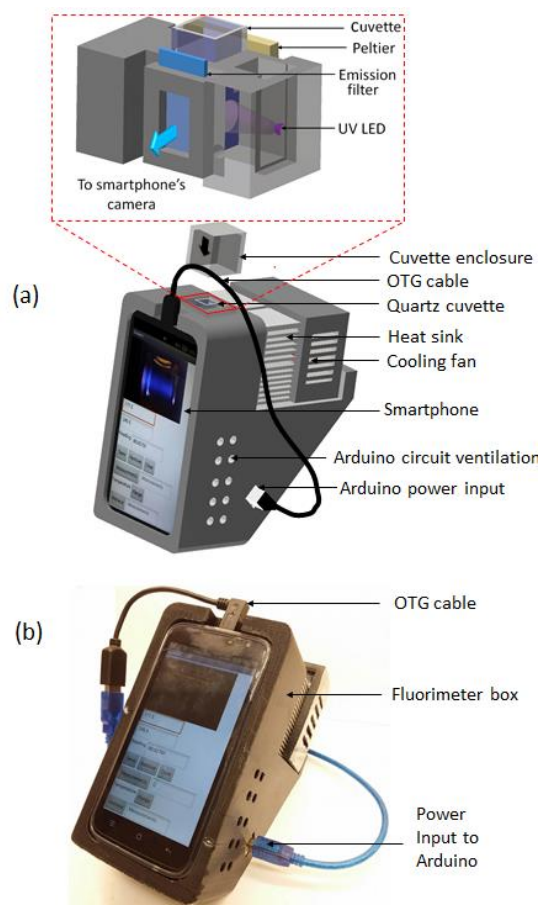

Fig. 2. 3D design of an Arduino-based smartphone fluorimeter. The complete 3D AutoCAD design of the fluorimeter (inset shows the detail structure of the sample chamber); and (b) the 3D printed device installed on an Android phone.

### Instrument Calibration

A step-by-step calibration of each of the individual components was carried out. This includes the optical stability of the excitation source at different thermal conditions, and response of the Peltier and $T$-sensors. Figure S1 (Electronic Supplementary Information) summarises the results of experiments to calibrate each individual component of the system. The results show that the output emission of the excitation source varies linearly with input current (Fig. S1a). This allows the user to control the optical power using a potentiometer, contrasting with many commercial spectrometers where this is done by adjusting the excitation slit width which sacrifices the spectral resolution. Despite this adjustment, the UV LED also shows good stability over time (Fig. S1b) and $T$ (Fig. S1c). The stability of the smartphone's battery power supply has been verified by monitoring $I$ and $V$ from the micro USB port at different current levels. Both $I$ and $V$ are found to be stable at 5.87 mA and 5.11 V respectively across a load $R \sim 83\ \Omega$ over time (Fig. S1d).

**Temperature calibration**

The response of the temperature sensor in the smartphone fluorimeter was calibrated against a standard $K$-type thermocouple (Fluke 50-Series II Model 52). To do this, the temperature of the sample cuvette, filled with deionised water, was monitored both with the smartphone fluorimeter and simultaneously using the thermocouple directly immersed into the water. By direct immersing into the sample, the thermocouple reads the actual temperature which will be compared to the surface temperature of the cuvette recorded by the IR temperature sensor. The bottom surface of the cuvette is coated with a non-transparent black paint so that the temperature sensor receives emission only from the cuvette surface and block emission from the surroundings through the transparent cuvette walls. Figure 3 shows the comparison between temperature readings on the smartphone fluorimeter ($T_{ir}$) and from the standard thermocouple ($T_{th}$). From the linear fit of the data, different correlations between the two systems are obtained for heating and cooling conditions. The different responses to heating and cooling reflect hysteresis in the Peltier response. Two empirical equations were obtained and uploaded to the app to correct for this $T$ response during measurements:

For heating ($T \geq 25\ ^oC$)
$$T_{ir} = 0.77\, T_{th} + 7.6 \qquad (01)$$

For cooling ($T < 25\ ^oC$)
$$T_{ir} = 1.26\, T_{th} - 8.5 \qquad (02)$$

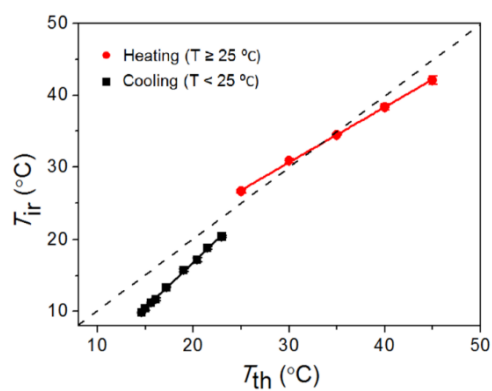

Fig. 3. Temperature response calibration of the smartphone fluorimeter IR temperature sensor with respect to a separate thermocouple. Raw data from IR temperature sensor are averaged after 3 times measurements and dashed line is the thermocouple response.

**Battery lifecycle**

The operational time of the complete system is estimated for a full discharge cycle of the smartphone battery (Blackview ZETA, $E$ = 2050 mAh). The total working time as well as actual $V$ and $I$ across some components were recorded under different loading conditions to the battery (Table S1 Electronic Supplementary Information). With minimum components connected, i.e. when only the excitation source, temperature sensor and Bluetooth module are ON, the smartphone fluorimeter can operate continuously and reliably for $t \sim 5$ h. Using a Peltier (2 x 2 cm$^2$), power consumption permitted a working time of $t \sim$ (60 - 70) mins for both heating and cooling. In lieu of future improved battery performance, adding a second battery can extend the device lifetime substantially but nonetheless this time was sufficient for the measurements reported here and sufficient to have a fully functioning, field portable fluorimeter that can perform as well as many benchtop instruments.

## Fluorescence measurements

The temperature response of steady-state fluorescence intensity ($F$) from two commonly available laser dyes, (Rhodamine B (RhB, **1**) and Rhodamine 6G (Rh6G, **2**)), a porphyrin emitter 5,10,15,20-tetraphenylporphyrin (**3**) and the chemosensor dye 6-(1,4,8,11-tetraazacyclotetradecane)2-ethyl-naphthalimide fluoro-ionophore (**4**) were measured to demonstrate the Arduino-controlled smartphone fluorimeter performance and capability.

The laser dyes RhB (**1**) and Rh6G (**2**) have a wide range of applications in chemical research whilst having opposite response with increasing temperature. RhB, for example, is often applied in non-contact sensing of $T$-changes produced by radiofrequency radiation, specifically in small biological samples.[35] By using two rhodamine dyes with opposite temperature effects, millimeter wave propagation inside a rectangular waveguide has been characterized.[36] This makes these dyes excellent test subjects for the fluorimeter. Figure 4ashows the decrease of fluorescence intensity in response to increasing temperature for the RhB solution in deionized water ([RhB] = 0.1 mM). Similar to RhB, in most cases an increase in $T$ results in a reduction in the fluorescence quantum yield because there is an increase in molecular collisions in solution and a rise in the amplitude of internal molecular vibrations, leading to higher non-radiative relaxation of the excited state and therefore greater fluorescence quenching. On the other hand, some organic molecules in aqueous solution form associated complexes (dimers, trimers, and so forth), the concentration of which increase with temperature. These complexes are better shielded and have higher yields producing increases in fluorescence, explaining the small increase in fluorescence of Rh6G in deionized water with increased temperature. Consistent with that, Fig. 4b shows the measured increase of Rh6G (0.2 mM) fluorescence with increasing temperature. The rate of fluorescence change can be obtained from the slopes in Fig. 4 and expresses in percentage as

$$\eta = \pm \left( \frac{(F-F_0)/F_0}{\Delta T} \right) \times 100 \qquad (03)$$

From the linear fits in Fig. 4(a) and (b), the rates of fluorescence change in RhB ($\lambda_{em}$ = 600 nm) and Rh6G ($\lambda_{em}$ = 550 nm) are recorded as $\eta_{RhB}$ = - 1.58% /°C and $\eta_{Rh6G}$ = 0.51% /°C respectively. These observed values reasonably found within the values reported in literature.[35,37] The significantly higher rate of change of fluorescence of RhB makes these very useful for different *T*-responsive applications.

Porphyrins are robust, conjugated ring systems, often important cofactors found in many biomolecules including chlorophyll and vitamin $B_{12}$. Using the smartphone fluorimeter, the fluorescence ($\lambda_{em}$ = 650 nm) of 50 μM 5,10,15,20-tetraphenylporphirin (TPP, **3**) in acetone was recorded over the *T* range 10 to 40 °C, showing a very slow decrease of fluorescence for TPP with increased *T* (Fig. 4c). The rate of decay is recorded as $\eta_{TPP}$ = - 0.45% /°C. The robustness along with rigidity compared to the dyes makes this porphyrin much less temperature dependent. Relatively high thermal stability of porphyrin dyes is also reported by others where no decomposition was observed until 200 °C.[38]

The temperature-responsive fluorescence ($\lambda_{em}$ = 450 nm) of a chemosensor dye **4** was also measured using the Arduino driven smartphone fluorimeter. Chemosensor dyes of this type are widely used to quantify metal ion (such as $H^+$, $Zn^{2+}$ and $Cu^{2+}$) concentrations in environmental and biological samples.[39-41] These systems can be easily adapted to smartphone devices to enable off-site rapid diagnosis and have been used to map pH around Sydney[14] as well as characterising metal ion concentrations including any unintentional metal ion disruption in water supply systems. Temperature effects can reverse their metal ion selectivity, making it possible to use fluorescence to monitor multiple metal ions in the one sample.[27] For example, it has been shown that, in the presence of electron donor $Cu^{2+}$, photo-induced electron transfer (PET) in the ligand **4** increases whereas net fluorescence, $\Delta F$ ($F-F_0$) decreases as a function of *T*, which is the opposite of the case with $Zn^{2+}$ ion.[27] However, the rate at which the ligand/complex bends or twists can be slowed by reducing the temperature of the solvent system, which means that at sufficiently low temperature (*T* ≤ 15 °C), the effect of $Zn^{2+}$ on the overall fluorescence emission is greatly diminished, effectively zero. Therefore, temperature can be used to discriminate between metal ions.

To demonstrate the potential when both $Zn^{2+}$ and $Cu^{2+}$ are present with the ligand **4** (5 μM in HEPES buffer, pH = 8.0) in a concentration ratio 1:1:1, a net decrease in fluorescence with decreasing temperature is observed (Fig. 4d). This net fluorescence decreases at *T* = 10 °C to the same value in the sample containing 1:1 $Zn^{2+}$:$Cu^{2+}$ as the $Cu^{2+}$ case alone,[27] indicating that the $Zn^{2+}$-bound species is not triggering a change in fluorescence output at this temperature - $Cu^{2+}$ preferentially binds with the ligand ahead of $Zn^{2+}$. Thus the presence of the $Cu^{2+}$ can be identified.

**Time-resolved relaxation measurements**

Our previous study on the chemosensor **4**[27] shows that, in the presence of $Zn^{2+}$ only, the ligand is bent and the fluorescence emission through intramolecular charge transfer is characterised by an intensity growth of up to 1 minute. This is a remarkably long timescale explained by a physical movement in the electronic distribution around the molecule. That such movement can be controlled by optical excitation led to the proposal of using twist- and bend-induced charge transfer to enable photonic powered molecular nanobots and nanobot technology.[28] Here, we used the smartphone fluorimeter to time-resolve the evolution of fluorescence measurements to determine this twisting and bending rate, identified through a bending rate constant. In this case, the smartphone app captures automatically a total of $n$ = 75 images of the fluorescent sample at a regular interval, $\delta t$ = 4 seconds. The fluorescence intensities were plotted locally on the smartphone screen as $F$ vs $t$ and also sent to a computer using the Internet.

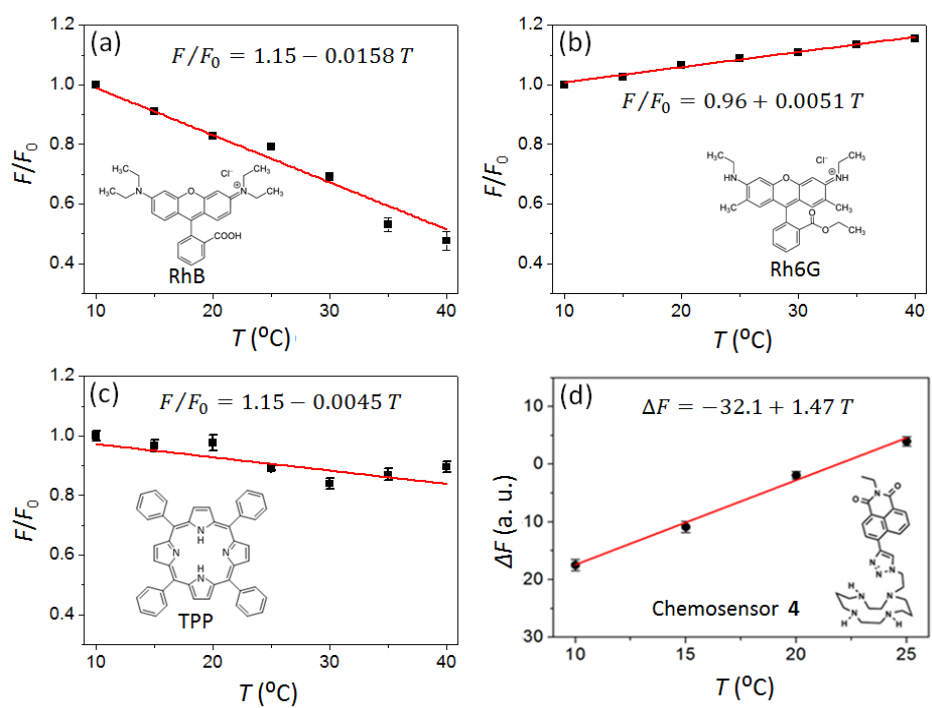

Fig. 4 Temperature dependent fluorescence (*F*), normalised by the initial fluorescence ($F_0$), measurements of two laser dyes– (a) rhodamine B; (b) rhodamine 6G; (c) a porphyrin emitter 5,10,15,20-tetraphenylporphirin. (d) The change of net fluorescnece, $\Delta F$ ($F-F_0$) at different *T* of a chemosensor dye (6-(1,4,8,11-tetraazacyclotetradecane)2-ethyl-naphthalimide) with dye:$Zn^{2+}$:$Cu^{2+}$ = 1:1:1 concentration ratio. Data were obtained using the Arduino driven smartphone fluorimeter. The error bars represent standard devaiation of 3 times measurements of each sample.

In order to understand the factors that impact on this mechanical bending, both the optical and thermal excitation using the smartphone fluorimeter were tuned and the rate of "bending and twisting" measured. Optical excitation was varied by adjusting the current through the UV LED (Fig. S1a shows linear relationship between optical output and diode current). Thermal excitation was varied by adjusting the temperature of the sample. The emission intensity is fixed to the peak wavelength of the fluorescent band, $\lambda \sim 450$ nm which as expected grows at a constant rate with time for all given excitation intensities. The normalised data in Fig. 5 (a plot of $F$ versus $t$ for different 370 nm diode currents) shows little variation with intensity within experimental error. The small, reproducible increase in intensity with increasing diode current is due to local heating of the components which is not able to be dissipated quickly enough. A reasonable conclusion is that there is no observable control of the rate of bending arising from optical intensity – this is unsurprising given that electronic excitation is fast (ns) whereas mechanical relaxation is slow and an indirect consequence of optical excitation. So fine-tuning this optical "nanobot"'s rate of movement using light is not immediately feasible.

To alter the rate of bending and/or twisting requires direct excitation of the mechanical process – the most obvious way to achieve this is by varying temperature (hinted by the data in Fig 5). Figure 6 shows the normalised intensity as a function of time for experiments at a range of different temperatures. The rate of fluorescence change increases with increasing temperature, as anticipated. Assuming that bending is the dominant process involved here, it is plausible to quantify the rate of molecular bending as a function of bend angle, directly from the emission data. The single exponential growth of emission can be expressed directly to represent the rate of bending as a function of $T$:

$$\theta_n = \theta_0(T)e^{-kt}$$

where $\theta_0$ is the initial angle, $k$ is the rate constant, $T$ is the temperature and $t$ is the time of exposure. By knowing the initial angle $\theta_0$, the bending angle $\theta_n$ ($n$ = 1, 2, 3…), as shown in Fig. 7, can be quantified at a given $T$. From the exponential fits of the normalized data, it is possible to extract the rate constant (min$^{-1}$) at each temperature used in Fig. 6 as well as for each current used in Fig. 5. These are summarised in Table 1. Values of the measured rate constant on the smartphone agree with those reported with a standard benchtop fluorimeter.[27] This result indicates that temperature tuning offers a tangible route to controlling useful mechanical parts and even full robotic functionality. This greatly enhances the potential for enabling molecular machines.[42,43] Future designs can seek to combine and optimise optically induced charge transfer and heating to make this more efficient.

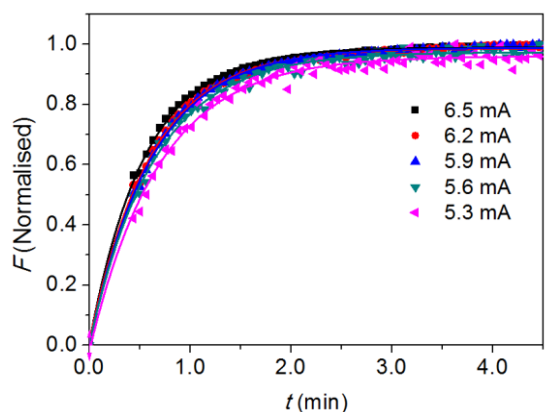

Fig 5. Emission intensity, $F$ of **4** (normalised to the final intensity) with Zn$^{2+}$ (1:1) at room temperature (22 °C) for different excitation current as a function of time, $t$. Measurements at each $I$ were performed for 3 different samples on the smartphone intensity fluorimeter and average data were plotted on a computer.

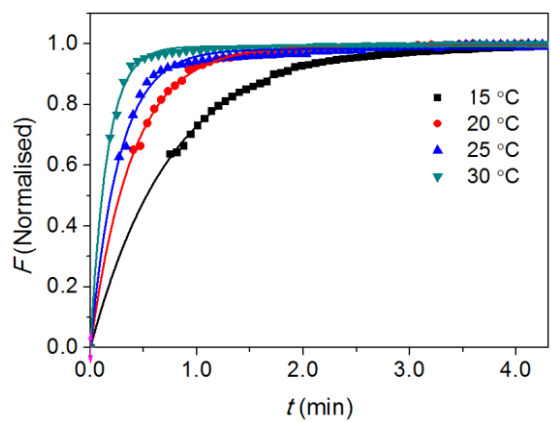

Fig 6. Emission intensity, $F$ of **4** with Zn$^{2+}$ (1:1) at four temperatures (15, 20, 25 and 30 °C) for fixed excitation at 370 nm as a function of time, $t$. Measurements at each $T$ were performed for 3 different samples on the smartphone intensity fluorimeter and average data were plotted on a computer.

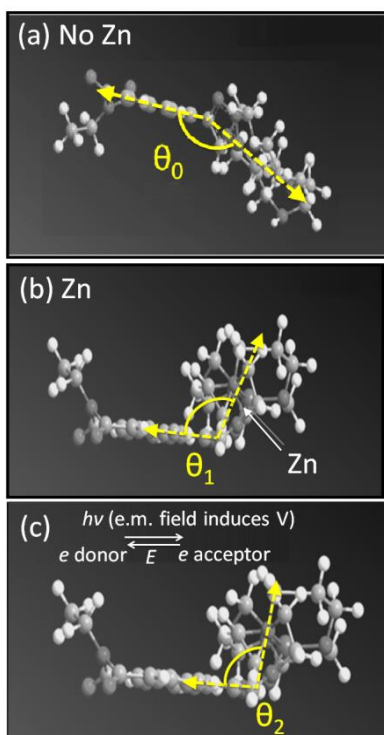

Fig 7. Simulation of **4** (a) without $Zn^{2+}$; (b) with $Zn^{2+}$ and (c) with $Zn^{2+}$ and optical excitation ($\lambda_{ex}$ = 370 nm). As a result of electrostatic interactions, the cyclam group is more twisted and bent out of plane with the naphthalimide group which increases with $Zn^{2+}$ and optical interaction.

Table 1: Bending rate constant ($k$) at different excitation current ($I$) and temperature ($T$).

| $T$ (°C) | $I$ (mA) | $k$/min$^{-1}$ |
|---|---|---|
| 22 | 5.3 | 1.45 |
| 22 | 5.6 | 1.55 |
| 22 | 5.9 | 1.61 |
| 22 | 6.2 | 1.64 |
| 22 | 6.5 | 1.81 |
| 15 | 5.9 | 1.94 |
| 20 | 5.9 | 2.57 |
| 25 | 5.9 | 2.84 |
| 30 | 5.9 | 3.64 |

## Conclusions

Both the steady-state and time-based evolution of fluorescence measurements at different temperatures have been demonstrated using a novel, self-contained, portable, smartphone intensity fluorimeter. There remains considerable room to improve the device's capabilities further. Extending the device to allow fluorescence spectra to be recorded can be readily achieved by introducing a dispersive element into the system.[18] In terms of improving the temporal resolution of the fluorescence measurement, the imaging camera can be replaced by a fast photodetector, the subject of ongoing work. Future technologies, such as smart optical chip screens,[44] optical chips[45] including quantum optical chips and quantum single photon emitters such as those based on self-assembled silica[46] to accelerate processing times, enable encryption and reduce power consumption, along with next generation batteries, will only enhance the device performance.

The actual performance of the device demonstrated equal, if not superior, performance to previously reported measurements using benchtop equipment.[27] Further, the temperature-responsive fluorescence of several widely used fluorescent dyes **1–4** has been measured using the smartphone fluorimeter and found to be consistent with reported values elsewhere. The ability to use this instrument to separate the different responses of different metal ions in solution has been practically demonstrated. Time-resolved fluorescence intensity emitted from a chemosensor dye **4** was characterised both in terms of varying optical excitation intensity and thermal excitation. In addition, we have extracted rate constants associated with twisting and bending of the dye and explored the potential for molecular mechanical device technology.

Overall, the wireless/Bluetooth combination enabled a compact portable system where data was transmitted to central computers. Whilst Bluetooth has its limitations, it is low cost and the principles equally apply to newer wireless technologies. The instrument clearly has the potential to enable centralized collection and analysis of data from many such instruments across a field, aided by wireless networking, remote analysis, and individual mapping in real-time. Analysis elsewhere can be sent back to all devices to further enhance capabilities in the field. In addition to applications in chemical and biological research, the instrument has significant potential to demonstrate important chemical and physical science in the field and in resource limited arenas. This in fact forms the foundation for the true democratisation of science, seeded by community IoT and 3D printing, bringing a future that extends it well beyond traditional borders of academic institutions, perhaps the single most remarkable and impactive aspect of this technology.

## Acknowledgements

Authors acknowledge funding from Australian Research Council (ARC) through the grant DP140100975 and DP120104035, and an Edmund Optics Education Award in 2015 and private funding from J. Canning. M. A. Hossain acknowledges an International Postgraduate Research Scholarship (IPRS) at The University of Sydney. Z. Yu acknowledges a summer scholarship at the School of Chemistry, The University of Sydney.

## References

1. K. E. McCracken and J.-Y. Yoon, *Anal. Methods*, 2016, **8**, 6591.
2. D. N. Breslauer, R. N. Maamari, N. A. Switz, W. A. Lam and D. A. Fletcher, *PLoS ONE*, 2009, **4**, e6320.
3. Z. J. Smith, K. Chu, A. R. Espenson, M Rahimzadeh, A Gryshuk, M. Molinaro, D. M. Dwyre, S. Lane, D. Matthews and S. W. Hogiu, *PLoS ONE*, 2011, **6**, e17150.

4. B. Y. Chang, *Bull. Korean Chem. Soc.*, 2012, **33**, 549.
5. M. E. Quimbar, K. M. Krenek, and A. R. Lippert, *Methods*, 2016, **109**, 123. (http://dx.doi.org/10.1016/j.ymeth.2016.05.017).
6. S. Dutta, G. P. Saikia, D. J. Sarma, K. Gupta, P. Das, P. Nath, *J. Biophotonics*, 2016, 1–11.
7. E. Petryayeva and W. R. Algar, *Anal. Bioanal. Chem.*, 2016, **408**, 2913.
8. S. Sumriddetchkajorn, K. Chaitavon, and Y. Intaravanne *Sens. and Actuators B*, 2014, **191**, 561.
9. L. J. Loh, G. C. Bandara, G. L. Weber and V. T. Remcho *Analyst*, 2015, 140, 5501.
10. N. S. K. Gunda, S. Naicker, S. Shinde, S. Kimbahune, S. Shrivastavac and S. Mitra, *Anal. Methods*, 2014, **6**, 6236.
11. T. S. Park and J. Y. Yoon, *IEEE Sensors J.*, 2015, **15**, 1902.
12. S. Yu, W. Xiao, Q. Fu, Z. Wu, C. Yao, H. Shen and Y. Tang, *Anal. Methods*, 2016, **8**, 6877.
13. C. A. D. Villiers, M. C. Lapsley and E. A. H. Hall, *Analyst*, 2015, **140**, 2644.
14. M. A. Hossain, J. Canning, S. Ast, P. J. Rutledge and A. Jamalipour, *Phot. Sensors*, 2015, **5**, 289.
15. M. A. Hossain, J. Canning, S. Ast, P. J. Rutledge, T. L. Yen and A. Jamalipour, *IEEE Sensors J.*, 2015, **15**, 5096.
16. J. Canning, A. Lau, M. Naqshbandi, I. Petermann and M. J. Crossley, *Sensors*, 2011, **11**, 7055.
17. K. Yang, H. P. Soroka, Y. Liu, and F. Lin, *Lab Chip*, 2016, **16**, 943.
18. M. A. Hossain, J. Canning, S. Ast, K. Cook, P. J. Rutledge and A. Jamalipour, *Opt. Lett.*, 2015, **40**, 1737.
19. S. Dutta, D. Sarma, A. Patel and P. Nath, *IEEE Phot. Tech. Lett.*, 2015, **27**, 2363.
20. Elise K. Grasse, Morgan H. Torcasio and Adam W. Smith, *J. Chem. Educ.*, 2016, **93**, 146.
21. M. A. Hossain, J. Canning, K. Cook, and A. Jamalipour, *Opt. Lett.*, 2016, **41**, 2237.
22. C. Zhang, G. Cheng, P. Edwards, M.-D. Zhou, S. Zheng, and Z. Liu, *Lab Chip*, 2016, **16**, 246.
23. L.-J. Wang, Y.-C. Chang, R. Sun, L. Li, *Biosen. Bioelectron*. 2017, **87**, 686.
24. H. Yu, Y. Tan and B. T. Cunningham, *Anal. Chem.*, 2014, **86**, 8805.
25. Q. Wei, H. Qi, W. Luo, D. Tseng, S. J. Ki, Z. Wan, Z. Göröcs, L. A. Bentolila, T.-T. Wu, R. Sun, and A. Ozcan, *ACS Nano*, 2013, **7**, 9147.
26. B. B. Das, Feng Liu and R R Alfano, *Rep. Prog. Phys.*, 1997, **60**, 227.
27. J. Canning, S. Ast, M. A. Hossain, H. Chan, P. J. Rutledge, and A. Jamalipour, *Opt. Mat. Exp.*, 2015, **5**, 2675.
28. J. Canning, *in Proc. Nonlinear Photonics 2016* (paper JT4A.2).
29. K. P. Carter, A. M. Young and A. E. Palmer, *Chem. Rev.*, 2014, **114**, 4564.
30. D. Erickson, D. Sinton, and D. Li, *Lab Chip*, 2003, **3**, 141.
31. D. Ross, M. Gaitan and L. E. Locascio, *Anal. Chem.*, 2001, **73**, 4117.
32. U. Seger-Sauli, M. Panayiotou, S. Schnydrig, M. Jordan, P. Renaud, *Electrophoresis* 2005, **26**, 2239.
33. Arduino Uno and Genuino Uno. Available online: https://www.arduino.cc/en/Main/ArduinoBoardUno
34. F. J. DiSalvo, *Science* 1999, **285**, 703.
35. Y. Y. Chen and A. W. Wood, Bioelectromagnetics, 2009, **30**, 583.
36. N. Kuzkova, O. Popenko, and A. Yakunov, *Int. J. Biomed. Imaging*, 2014, **2014**, 243564.
37. V. K. Natrajan and K. T. Christensen, *in the proc. 14th Int Symp on Applications of Laser Techniques to Fluid Mechanics Lisbon, Portugal, July, 2008.*
38. C. Zhuang, X. Tang, D. Wang, A. Xia, W. Lian, Y. Shi and T. Shi *J. Serb. Chem. Soc.*, 2009, **74**, 1097.
39. Y. H. Lau, J. R. Price, M. H. Todd, and P. J. Rutledge, *Chem. Eur. J.*, 2011, **17**, 2850.
40. S. Ast, S. Kuke, P. J. Rutledge and M. H. Todd, *Eur. J. Inorg. Chem.*, 2015, **1**, 58.
41. E. Tamanini, A. Katewa, L. M. Sedger, M. H. Todd, and M. Watkinson, *Inorg. Chem*., 2009, **48**, 319.
42. P. Mobian, J. -M. Kern, J. -P. Sauvage, *Angew. Chem. Int. Ed.*, 2004, **43**, 2392.
43. https://www.nobelprize.org/nobel_prizes/chemistry/laureates/2016/advanced-chemistryprize2016.pdf
44. J. Lapointe, M. Gagné, M.-J. Li, and R. Kashyap, *Opt. Express*, 2014, **22**, 15473.
45. CNW team "Generic Optical Chip Brings Down Chip Costs," Chipsnwaters Jan 29, 2016 Available online: http://chipsnwafers.electronicsforu.com/2016/01/29/generic-optical-chip-brings-down-chip-costs/
46. M. Naqshbandi, J. Canning, B. C. Gibson, M. M. Nash, M. J. Crossley, *Nat. Commun.,* 2012, **3**, 1188.

# Electronic Supplementary information

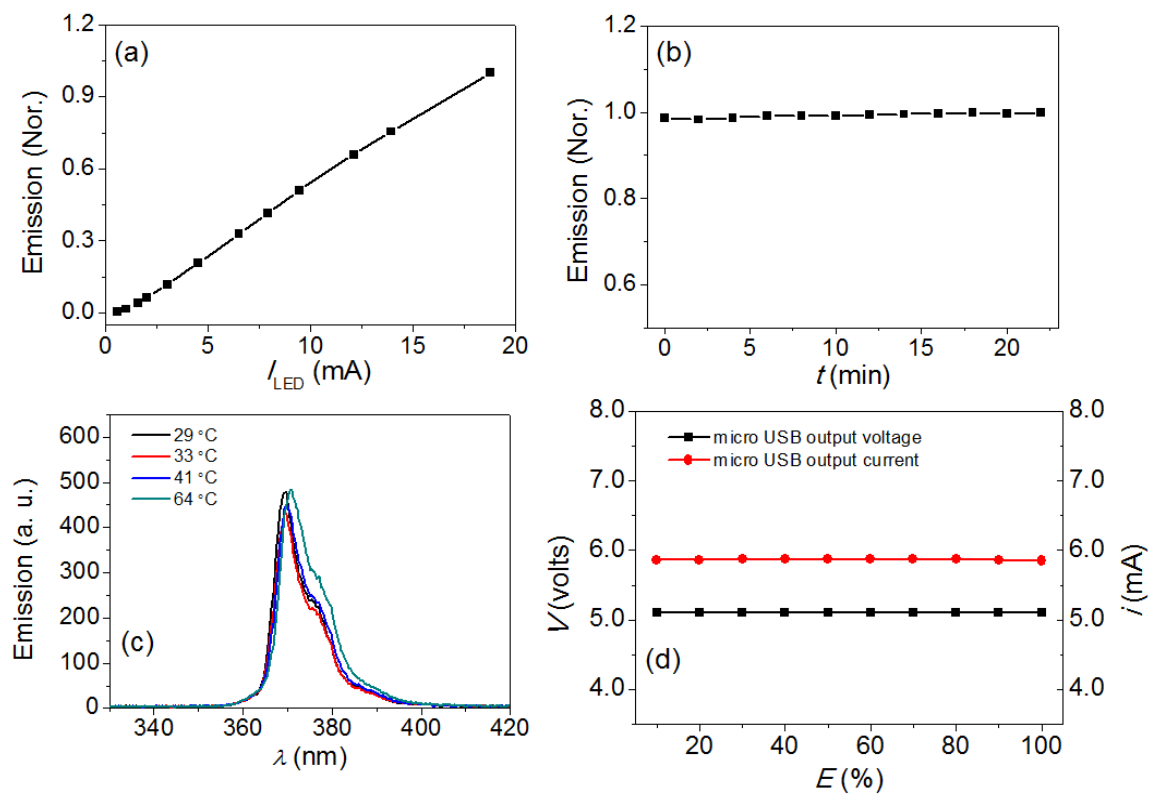

**Fig. S1**. Characterisation results of the excitation source and power supply of the smartphone fluorimeter. (a) Linearity of optical emission at different level of diode current ($I_{LED}$); (b) stability of optical emission over the time ($t$); (c) thermal stability of the excitation source emission spectrum; and (d) stability of the supplied voltage ($V$) and current over the full discharging cycle of the smartphone battery energy ($E$).

**Table S1**: Voltage ($V$) and current ($I$) measurements across all electronics components at different load connections. Minimum load: when only excitation source (UV LED), temperature sensor and Bluetooth are connected to the Arduino module. Maximum load: When a 2 x 2 cm Peltier was added with the minimum load setting. The UV LED is failed to keep ON due to the voltage regulation in the Arduino circuit.

|  | Peltier | | Fan | | LED | | Bluetooth | |
|---|---|---|---|---|---|---|---|---|
|  | $V$ (volts) | $I$ (A) | $V$ (volts) | $I$ (mA) | $V$ (volts) | $I$ (mA) | $V$ (volts) | $I$ (mA) |
| **Minimum load** (LED, temperature sensor and Bluetooth module) | - | - | - | - | 3.43 | 5.5 | 5 | - |
| **Maximum load** 2 x 2 cm Peltier, Cooling fan, LED, temperature sensor and Bluetooth module) | 0.92 | 0.41 | 2.83 | - | 3.01 | 0 (off) | 2.83 | - |